\documentclass{article}
\usepackage{template,amsmath,graphicx}
\usepackage{booktabs,amssymb}
\usepackage{cite}

\usepackage{algorithm}
\usepackage{algorithmic} 
\title{PUSHING THE LIMITS OF SELF-SUPERVISED SPEAKER VERIFICATION USING REGULARIZED DISTILLATION FRAMEWORK}
\name{Yafeng Chen, Siqi Zheng, Hui Wang, Luyao Cheng, Qian Chen}
\address{Speech Lab of DAMO Academy, Alibaba Group \\
\tt \normalsize\{chenyafeng.cyf, zsq174630\}@alibaba-inc.com}

\begin{document}

\maketitle
\begin{abstract}
Training robust speaker verification systems without speaker labels has long been a challenging task. Previous studies observed a large performance gap between self-supervised and fully supervised methods. In this paper, we apply a non-contrastive self-supervised learning framework called DIstillation with NO labels (DINO) and propose two regularization terms applied to embeddings in DINO. One regularization term guarantees the diversity of the embeddings, while the other regularization term decorrelates the variables of each embedding. The effectiveness of various data augmentation techniques are explored, on both time and frequency domain. A range of experiments conducted on the VoxCeleb datasets demonstrate the superiority of the regularized DINO framework in speaker verification. Our method achieves the state-of-the-art speaker verification performance under a single-stage self-supervised setting on VoxCeleb. Code has been made publicly available at https://github.com/alibaba-damo-academy/3D-Speaker.
\end{abstract}
\vspace{-0.1cm}
\section{Introduction}
\label{sec:intro}

Deep learning methods have achieved significant performance gains on speaker verification (SV) task. The availability of large labeled datasets and data augmentation methods have spurred remarkable improvements. However, collecting large amounts of labeled data in the real world is laborious and expensive. Therefore, self-supervised learning (SSL), that requires only unlabeled data, provides an alternative solution for learning representations from speech. 

Different unsupervised and self-supervised frameworks have been proposed to obtain meaningful speaker representations as in \cite{XiaZWYY21, CaiWL21, cho2021jhu, TaoLDHL22, JungKHLKC22, heo2022self, SangLLAW22, HanCQ22, ZhengLSL19a, ZhengLSL19}. More recently, some self-supervised SV systems propose two-stage SSL training \cite{CaiWL21, TaoLDHL22, HanCQ22}. In stage I, a speaker encoder is trained to obtain speech representations in a completely self-supervised manner. In stage II, a clustering algorithm is adopted to generate pseudo-labels for each utterance based on the previous learned representations. Then a new model is trained based on the estimated pseudo-labels iteratively in supervised learning. However, the second stage requires a ``good enough" estimate of the number of speakers in the training data, such as $M$ = 6000 in \cite{CaiWL21, TaoLDHL22} when using the development portions of VoxCeleb2 \cite{ChungNZ18}. Inaccurate settings will lead to significant performance degradation. This prior assumption restrains us from training SSL models from numerous unlabeled speech data involving a large but unknown number of speakers, which is self-contradictory with the original purpose of using SSL. Therefore, we stick to the single-stage approaches in self-supervised SV, without making assumptions on the overall distribution of training data.

Single-stage methods can be roughly divided into contrastive and non-contrastive ones. Xia et al. \cite{XiaZWYY21} applies SimCLR framework \cite{ChenK0H20} to extract robust speaker embeddings by directly maximizing the similarity between augmented positive pairs and minimizing the similarity of negative pairs via a contrastive InfoNCE loss \cite{abs-1807-03748, ZhengSC22}. Such frameworks require large number of negative samples and huge batch size during training. He et al. tried to remove the batch size limitations by introducing Momentum Contrast (MoCo) framework \cite{He0WXG20}. MoCo uses a dynamic queue to store negative samples and compute the InfoNCE loss. 

Most contrastive methods either require a huge batch size or maintain a dynamic queue, both of which are computationally expensive. Furthermore, contrastive methods in speaker verification select negative samples through random sampling, which may result in false negative samples. It will cause instability in network training and degrade system performance since the contrastive loss may push the potential positive samples further to each other.

Non-contrastive methods are free of such worries because the negative samples are not required in the training process. In fact, these methods \cite{cho2021jhu, JungKHLKC22, heo2022self, SangLLAW22, HanCQ22} have shown comparable or better performance compared to contrastive counterparts. Among non-contrastive frameworks, BYOL \cite{GrillSATRBDPGAP20} and DINO \cite{CaronTMJMBJ21} are the more attractive ones proposed in computer vision. BYOL is composed of online and target networks. The online network predicts the target network representation of the same utterance under a different augmented view. In \cite{SangLLAW22}, a self-supervised regularization term is proposed in speaker verification inspired by BYOL. DINO is a self-distillation framework which contains a teacher and a student network with an identical architecture but different parameters. It compares multiple different views generated from a single utterance and employs a self-distillation method which is widely used in speaker verification \cite{cho2021jhu, JungKHLKC22, heo2022self, HanCQ22}.

Due to the lack of negative samples, model collapse has been a common problem for non-contrastive methods. The network is more inclined to map positive pairs to the same or similar positions in the embedding space, resulting in trivial solutions. To avoid model collapse, \cite{SangLLAW22} introduces the regularization MLP structure which is applied on online network. DINO applies sharpening and centering techniques to the teacher output and uses exponential moving average (EMA) update strategy \cite{heo2022self}. 

Inspired by \cite{BardesPL22}, we propose two regularization terms applied to embeddings to further alleviate this problem in speaker verification task. One regularization term guarantees the diversity of the embeddings within a batch, while the other regularization term decorrelates the variables of each embedding. In order to effectively capture the utterance-dependent variability into the embedding, three kinds of augmentation strategies - WavAugment \cite{SnyderGSPK18}, SpecAugment \cite{ParkCZCZCL19} and Acoustic feature shuffling \cite{LiFCGSD22} are investigated. Experimental results indicate that the proposed Regularized DINO framework (RDINO) can significantly boost the performance of self-supervised speaker verification.


\section{DINO applied in SV}
\label{sec:format}

Inspired by \cite{CaronTMJMBJ21}, we apply DINO to speaker verification. DINO is a self-distillation framework where the outputs of a teacher network are used as ground truth to optimize a student network in parallel. Various types of cropped and augmented views are constructed from an utterance, divided into local and global views depending on the length of segments. Global views go through teacher network and all views are fed into student network. The global information learned by teacher guides the training of student, therefore encouraging "local-to-global" correspondences. Both networks share the same architecture \emph{g} with different sets of parameters. 

Take teacher module as example, it is composed of a backbone \emph{f} (ECAPA-TDNN), and of a projection head $h: g = f \circ h$. The speaker embedding is the backbone $f$ output. ECAPA-TDNN is the most commonly used network structure in speaker verification, with powerful feature extraction capability. The projection head $h$ consists of three fully connected (FC) layers with hidden dimension 2048-2048-256 followed by $L2$ normalization and a weight normalized FC layer with $K$ dimensions. Cross-entropy loss is calculated to minimize the probability distribution as follows.

\begin{equation}
  L_{CE} = \sum_{\textbf{x} \in \textbf{X}_{l}}\sum_{\substack{\textbf{x}^{\prime} \in \textbf{X}_{l} \cup \textbf{X}_{s} \\ \textbf{x}^{\prime} \neq \textbf{x}}} H(P^{tea}(\textbf{x})| P^{stu}(\textbf{x}^{\prime}))
\end{equation}
where $H(a | b)=-a*log b$ is cross-entropy, $\textbf{X}_l = \{\textbf{x}_{l_1}, \textbf{x}_{l_2}\}$ stands for two long segments and $\textbf{X}_s = \{\textbf{x}_{s_1}, \textbf{x}_{s_2}, $
$\textbf{x}_{s_3}, \textbf{x}_{s_4}\}$ stands for four short segments. $P^{tea}$, $P^{stu}$ represent the output probability distributions of teacher network $g_{\theta}^{tea}$ and student network $g_{\theta}^{stu}$ respectively. Mathematically,
\begin{equation}
    P^{tea}(\textbf{x})=Softmax\left(g_{\theta}^{tea}(\textbf{x}) / \tau_t\right)
\end{equation}
$\tau_t$ is a temperature parameter that controls the sharpness of the teacher's output distribution, and a similar formula holds for $P^{stu}$ with temperature $\tau_s$.

To avoid model collapse, teacher network is updated by EMA of the student's parameters. In addition, sharpening and centering techniques are applied to the teacher output. Beyond that, we add diversity regularization and redundancy elimination regularization to conquer the model collapse problem, which will be introduced in the following section.

\begin{figure*}[hbt]
  \centering
  \includegraphics[scale=0.55]{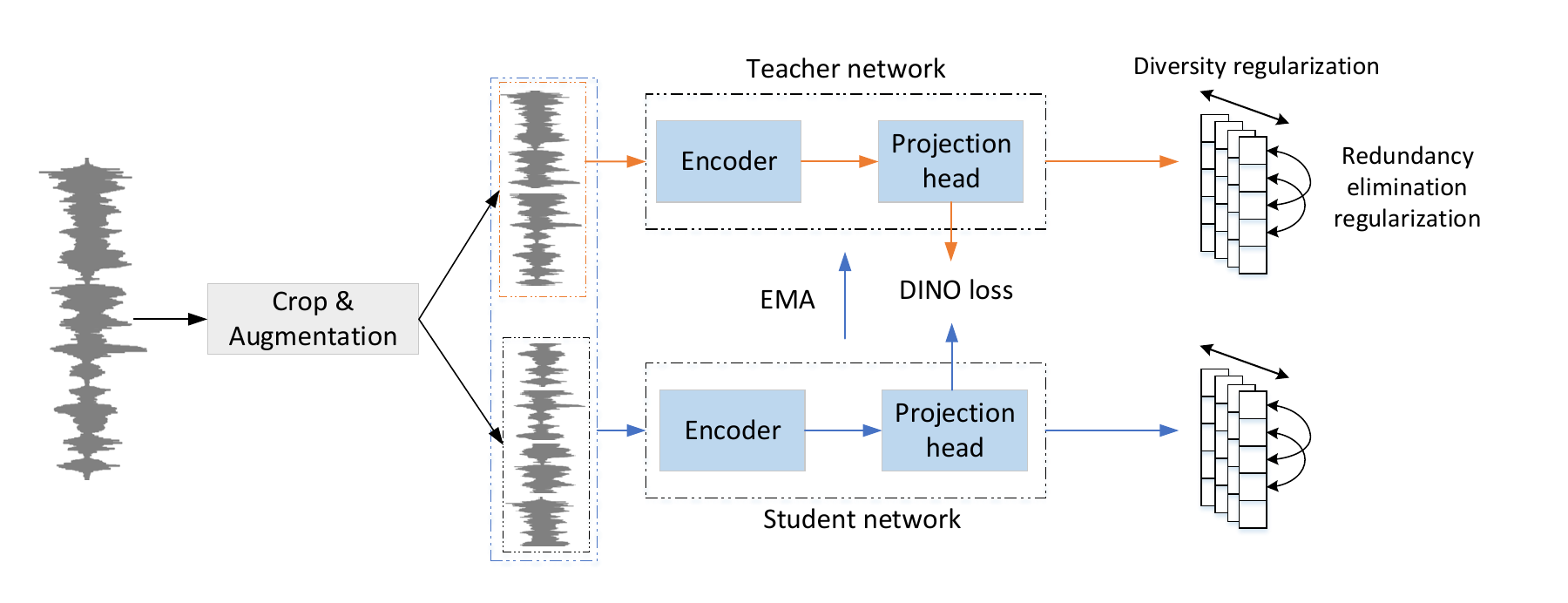}
  \caption{Overview of the proposed regularized DINO framework.}
\end{figure*}

\section{Proposed Method}
\label{sec:pagestyle}

In order to further alleviate the model collapse problem and improve the robustness of the speaker embeddings in DINO framework, we propose two regularization terms called diversity regularization and redundancy elimination regularization.

\vspace{-0.1cm}
\subsection{Diversity regularization}
\label{sssec:subhead}

The diversity regularization term guarantees the diversity of the embeddings within a batch. It forces the embeddings of utterances to be different and prevents trivial solutions. The diversity regularization loss can be calculated as shown in Eq. 3.  $\textbf{z}_{j}^{tea}$ and $\textbf{z}_{j}^{stu}$ stand for the teacher's and student's embeddings of dimension $j$ respectively. We calculate the standard deviation of each dimension in embeddings within one batch. $\epsilon$ is a small scalar preventing numerical instabilities.

\vspace{-0.2cm}
\begin{equation}
\begin{aligned}
  L_{DR} = &\frac{1}{d} \sum_{j=1}^d \max \left(0,1-\sqrt{\operatorname{Var}(\textbf{z}_j^{tea})+\epsilon}\right) \quad +\\
  &\frac{1}{d} \sum_{j=1}^d \max \left(0,1-\sqrt{\operatorname{Var}(\textbf{z}_j^{stu})+\epsilon}\right)
\end{aligned}
\end{equation}

\vspace{-0.2cm}
\subsection{Redundancy elimination regularization}
\label{sssec:subhead}
The redundancy elimination regularization term decorrelates the variables of each embedding while minimizing the redundancy. It attracts the covariances of each dimension in all speaker embeddings within a batch towards zero and prevents an informational collapse in which the variables would be highly correlated. The redundancy elimination regularization loss is calculated as follows.
\begin{equation}
  L_{RER} = \sum_{i}\sum_{j \neq i} C_{ij}^2
\end{equation}
where $C$ is the cross-correlation matrix computed between the global outputs of teacher network and student network along the batch dimension, and $C_{ij}$ is defined as Eq. 5:
\begin{equation}
C_{i j} = \frac{\sum_b z_{b, i}^{tea} z_{b, j}^{stu}}{\sqrt{\sum_b\left(z_{b, i}^{tea}\right)^2} \sqrt{\sum_b\left(z_{b, j}^{stu}\right)^2}}
\end{equation}
%
where $b$ indexes batch samples and $i,j$ index the embedding dimension. The redundancy elimination term decorrelates the different vector components of the embeddings by trying to equate the off-diagonal elements of the cross-correlation matrix to 0.

\subsection{Regularized DINO framework}
\label{sssec:subhead}

The overview of regularized DINO framework is depicted in Fig. 1.
First of all, we sample two long segments $\textbf{X}_l = \{\textbf{x}_{l_1}, \textbf{x}_{l_2}\}$ and four short segments $\textbf{X}_s = \{\textbf{x}_{s_1}, \textbf{x}_{s_2}, \textbf{x}_{s_3}, \textbf{x}_{s_4}\}$ from an utterance with a multi-crop strategy and different data augmentation strategies. Then $\textbf{X}_l$ are first encoded by $f^{tea}_{\vartheta}$ and $f^{stu}_{\vartheta}$ respectively into their representations regarded as speaker embeddings $\textbf{E}^{tea}_{l}$ and $\textbf{E}^{stu}_{l}$. Then $\textbf{E}^{tea}_l$ and $\textbf{E}^{stu}_l$ are mapped by four FC layers with hidden dimension $2048, 2048, 8192, 256$ and a weight normalized FC layer with $65536$ dimensions. We denote $\textbf{Z}_l^{tea} = [\textbf{z}_{l_1}^{tea}, \textbf{z}_{l_2}^{tea}]$ and $\textbf{Z}_l^{stu} = [\textbf{z}_{l_1}^{stu}, \textbf{z}_{l_2}^{stu}]$ as the teacher's output and student's output of dimension $8192$. The diversity regularization and redundancy elimination regularization loss are computed in $\textbf{Z} = \textbf{Z}_l^{tea} \cup \textbf{Z}_l^{stu}$ shown as Eq. 3 and Eq. 4. The speaker embedding network is jointly trained with the $L_{CE}$, $L_{DR}$ and $L_{RER}$. The overall loss is calculated as Eq. 6, the hyperparameter $\lambda$ controls the balance of all losses.
\begin{equation}
  Loss = L_{CE} + \lambda(L_{DR} + L_{RER})
\end{equation}
%

\section{Experiments and analysis}
\label{sec:typestyle}

\subsection{Experimental settings}
\label{ssec:subhead}

\subsubsection{Datasets and evaluation metrics}
\label{sssec:subsubhead}

To investigate the effectiveness of the proposed method, we conduct experiments on the VoxCeleb datasets. The development portions of VoxCeleb2 \cite{ChungNZ18} are used for training. It comprises 1,092,009 utterances among 5,994 speakers. Performance of all systems are evaluated on the test set of VoxCeleb1 \cite{NagraniCZ17}. No speaker labels are used during training in all experiments.

The results are reported in terms of two metrics, namely, the equal error rate (EER) and the minimum of the normalized detection cost function (MinDCF) with the settings of $P_{target}$ = 0.05 and $C_{fa} = C_{miss}$ = 1.

\subsubsection{Input features}
\label{sssec:subsubhead}
For each utterance, we use the multi-crop strategy for RDINO training in which 4s segments and 2s segments regarded as global views and local views respectively. The acoustic features used in the experiments are 80-dimensional Filter Bank (FBank) with 25ms windows and 10ms shift. Speech activity detection (SAD) is not performed as training data consists mostly of continuous speech. Mean and variance normalization is applied using instance normalization on FBank features.

\subsection{Data augmentation}
\label{sssec:subhead}
Data augmentation has been proven to be crucial for both supervised and self-supervised representation learning. Therefore, we explore the impact of data augmentation in our regularized DINO framework. WavAugment, SpecAugment and Acoustic feature shuffling are considered.

\noindent\textbf{WavAugment:} MUSAN corpus with SNR between 0 to 15 for additive noise and Room Impulse Response (RIR) for reverberation are applied to each long segment and short segment randomly.

\noindent\textbf{SpecAugment:} Apply time masking and frequency masking at acoustic feature level. The time masking length is 0 to 15 frames and frequency masking length is 0 to 6 dimensions. Each time we apply one randomly selected time mask and frequency mask on the FBank features.

\noindent\textbf{Acoustic feature shuffling:} To learn the sequential order invariant speaker embeddings, we shuffle the time order of acoustic feature frames at the segment scale of 50 frames. Detailed steps are discribed in \cite{LiFCGSD22}.

\subsection{Model configurations and implementation details}
\label{sssec:subhead}
We exploit the ECAPA-TDNN with attentive statistical pooling as the encoder model $f$, followed by a 512-d FC layer. The projection head $h$ consists of four FC layers with hidden size of $2048, 2048, 8192, 256$ and a weight normalized FC layer with 65536 dimensions. We train the model 60 epochs using the stochastic gradient descent (SGD) optimizer with momentum of 0.9. The learning rate is linearly ramped up during the first 10 epochs to 0.325. After this warmup, we decay the learning rate with a cosine schedule. The temperature $\tau_t$ and $\tau_s$ are set to 0.04 and 0.1 respectively.

\subsection{Results and analysis}
\label{sssec:subhead}
We investigate the performance of regularized DINO framework and evaluate it on the VoxCeleb1 test set. We compare our method to \cite{XiaZWYY21, SangLLAW22, cho2021jhu, HanCQ22, heo2022self, huh2020augmentation, abs-2010-11433, TaoLDHL22} which are recently proposed self-supervised learning architectures as shown in Table 1.

\begin{table}[h]
    \caption{Comparison with self-supervised learning models}
    \vspace{0.2cm}
    \centering
    \begin{tabular}{p{4cm}p{1.5cm}p{1.5cm}}
    \toprule
    \textbf{Framework} & \textbf{EER(\%)} & \textbf{MinDCF} \\
    \midrule
    AP+AAT \cite{huh2020augmentation} & 8.65 & 0.454 \\
    MoCo + WavAug \cite{XiaZWYY21} & 8.23 & 0.590 \\
    CEL \cite{abs-2010-11433} & 8.01 & N / R \\
    Contrastive \cite{TaoLDHL22} & 7.36 & N / R \\
    SSReg \cite{SangLLAW22} & 6.99 & 0.434 \\
    DINO + Cosine loss \cite{HanCQ22} & 6.16 & 0.524 \\
    DINO \cite{cho2021jhu} & 4.83 & N / R \\
    DINO + CL \cite{heo2022self} & 4.47 & 0.306 \\
    \midrule
    \textbf{RDINO \ (Ours)} & \textbf{3.29} & \textbf{0.247} \\
    
    \bottomrule
    \end{tabular}

\end{table}

It can be observed that the RDINO system decreases the EER to 3.29\%, which outperforms the previous self-supervised methods (4.47\% EER) by a significant +26.4\% relative improvement. This finding verifies the effectiveness of the two regularization terms.

\begin{table}[h]
    \caption{The effect of the weight $\lambda$ in RDINO}
    \vspace{0.2cm}
    \centering
    \begin{tabular}{p{1.8cm}p{1.8cm}p{1.8cm}}
    \toprule
    \textbf{Weight} & \textbf{EER(\%)} & \textbf{MinDCF} \\
    \midrule
    $\lambda$ = 0 & 3.62 & 0.262 \\
    $\lambda$ = 0.1 & 3.40 & 0.259 \\
    \textbf{$\lambda$ = 0.2} & \textbf{3.24} & \textbf{0.252} \\
    \textbf{$\lambda$ = 0.3} & \textbf{3.29} & \textbf{0.247} \\
    $\lambda$ = 0.4 & 3.37 & 0.253 \\
    $\lambda$ = 0.5 & 3.52 & 0.251 \\
    
    \bottomrule
    \end{tabular}

\end{table}

Moreover, we also conduct experiments to investigate the effect of weight $\lambda$ in RDINO system as shown in Table 2. We observe that applying two regularization terms outperforms the DINO system with even a small weight $\lambda$. It achieves 3.24\% EER with $\lambda = 0.2$ and 0.247 MinDCF with $\lambda = 0.3$. The results show superiority of the proposed method in self-supervised speaker verification task.

Additionally, we study the impact of different data augmentation strategies including WavAugment, SpecAugment and Acoustic feature shuffling on the training data. The experimental results are as follows.

\vspace{-0.2cm}
\begin{table}[h]
    \caption{The impact of data augmentation in RDINO}
    \vspace{0.2cm}
    \centering
    \begin{tabular}{p{4cm}p{1.5cm}p{1.5cm}}
    \toprule
    \textbf{Augmentation} & \textbf{EER(\%)} & \textbf{MinDCF} \\
    \midrule
    No Augment & 20.9 & 0.777 \\
    WavAugment & 3.29 & 0.247 \\
    \ + SpecAugment & 5.35 & 0.369 \\
    \ + Acoustic feature shuffling & 3.45 & 0.250 \\
    
   \bottomrule
    \end{tabular}

\end{table}

It is observed that WavAugment is the most efficient strategy in the RDINO framework. If no augmentation applied in training process, the whole network was difficult to converge due to DINO's inherent property. We find that SpecAugment worsened the verification results, which is different from our empirical observation for supervised speaker verification. SpecAugment uses erasing operation on the acoustic feature level to improve the model generalization. But the Voxceleb1 test data may not contain enough variabilities in the spectral domain. Acoustic feature shuffling strategy generates no gain in RDINO system, which may be caused by the operation in short length of the local view.

\section{Conclusions}
\label{sec:pagestyle}

In this paper, we introduce the DINO framework with different augmentation strategies for self-supervised speaker verification. In order to address the model collapse problem and further improve the system performance, we propose two regularization terms in DINO which achieve excellent performance over the conventional self-supervised models. 

\bibliographystyle{IEEEbib}
\bibliography{template}

\end{document}